\documentclass{aa}
\usepackage{graphicx}
\usepackage{natbib}
\bibpunct{(}{)}{;}{a}{}{,}

\def\ebv{\mbox{$\rm E(B-V)$}}

\def\ms{\mbox{$\rm M_\odot$}}
\def\ds{\mbox{$\rm d_\odot$}}

\def\jj{\mbox{$\rm J$}}

\def\ks{\mbox{$\rm K_S$}}

\def\mevol{\mbox{$\rm m_{evol}$}}
\def\mobs{\mbox{$\rm m_{obs}$}}

\def\mtot{\mbox{$\rm m_{tot}$}}

\def\kms{\mbox{$\rm km\,s^{-1}$}}

\begin{document}

\title{Structure and stellar content analysis of the 
open cluster M\,11 with 2MASS photometry}

\titlerunning{M\,11 structure and stellar content}

\author{J.F.C. Santos Jr.\inst{1}, C. Bonatto\inst{2} \and E. Bica\inst{2}}

\institute{Departamento de F\'{\i}sica, ICEx, UFMG, CP\,702, 30123-970 Belo 
Horizonte, MG, Brazil \and Departamento de Astronomia, IF-UFRGS, CP\,15051,
91501-970 Porto Alegre, RS, Brazil}

\offprints{J.F.C. Santos Jr., \email{jsantos@fisica.ufmg.br}}

\date{Received / Accepted }

\abstract{An overall analysis of the structure and stellar content of M\,11 is presented, 
thanks to the wide-angle 2MASS spatial coverage. We derive photometric and structural 
parameters and discuss the spatial dependance of the luminosity and mass functions. 
Photometric parameters basically agree with previous ones mostly based on the optical.
 We obtained a core radius of 1.23\,pc
and a tidal radius of 29\,pc. In particular, the cluster is populous enough so that
the tidal radius could be obtained by fitting the three-parameter King profile
to the radial distribution of stars.
We analyzed the spatial distribution of mass functions, finding that the the slope changes
from $-0.73$ in the core  to $+2.88$ in the outer halo. The spatial distribution of
mass function slopes derived from 2MASS agrees with that derived from 
optical CCD data, which further confirms the reliability of 2MASS data for future analyses 
of this kind at comparable observational limits. We detect mass segregation up to
distances from the center of $\approx20\arcmin$. We emphasize that the mass 
function slope in the core is flatter than anywhere else as a consequence
of mass segregation. The derived total cluster mass is $\approx11\,000\,\ms$.

}

\maketitle

\section{Introduction}

The open cluster M\,11 (NGC\,6705, Mel\,213, Cr\,391, OCl\,76 - \cite{alt70}) 
is a concentrated, populous stellar 
system projected on the Scutum Cloud towards the central part
of the Galactic disk [$\rm\alpha(2000)=18h51m05s$, $\rm\delta(2000)=-6^\circ16\arcmin01\arcsec$,
$l=27.30^\circ$, $b=-2.77^\circ$]. Although dark clouds permeate the 
sky on the cluster direction, it is situated in a clear
area characterized by a relatively low interestelar extintion nearby the
Sagittarius arm. Were it a low surface brightness cluster  it would be
probably missed by surveys due to the rich field from the Galactic background 
stars. For several reasons, M\,11 has captured attention over the years,
not only for its intrinsic properties, but also for its contribution 
to the understanding of chemical and dynamical Galactic evolution.   

Being closer to the Galactic center than the solar radius, M\,11 
suffers from relatively 
stronger tidal effects, as well as more frequent interactions with molecular
clouds. 
The WEBDA database \citep{mer96} 
provides a distance from the sun $\ds=1877$\,pc,
reddening $\ebv=0.426$, apparent distance modulus $\rm (V-M_V)=12.69$,
age t$=200$\,Myr and metallicity $\rm [Fe/H]=0.13$. 

\cite{ms77} studied proper motions in M\,11 and obtained a velocity dispersion
$\rm\sigma_v=2.9\,\kms$, and an 
observed mass of $\approx{3000}$\,M$_{\odot}$.
The inner cluster region may have isotropic orbits, while the orbits in
the outer parts are probably eccentric with larger velocities in the radial
direction. 

\cite{m84} carried out a comprehensive analysis of M\,11 based on
proper motion and membership probability data (from \citealt{mps77})
as well as photographic photometry reaching $\rm V=20$ and $\rm B=21$.
By studying the cluster luminosity function out to a radius of 10', 
evidence was found that inside 2' the luminosity function is flatter
than for the outer region, implying mass segregation. The total 
observed mass estimated inside the radius of 10'
and considering stellar masses down to $0.7\,\ms$ 
was $4671\,\ms$. 

\cite{sbd90} inferred the cluster overall mass function (MF) slope $1<\chi<2.4$
using a population synthesis method and the integrated spectrum aided by
the HR diagram. The cluster visible light is dominated by the upper main
sequence and turnoff stars (B6-A2).

\cite{nspm02} studied the  spatial structure of a large sample of 
open clusters using photometric data from the DSS. For M\,11 they 
derived a core radius of $0.72\pm0.10$\,pc, in agreement with the one
obtained by \cite{m84}.

Recently, \cite{bb05}, \cite{bbs04},  \cite{bbd04}, \cite{bb03} and references therein 
undertook a systematic
study of open cluster parameters, structure and other fundamental properties
employing 2MASS photometry, making use of a spatial coverage as large 
as necessary for each case. For M\,11, a deep UBVRI CCD study was carried out
by  \cite{sbl99}, including a spatial dependance of the MF. We intend
to compare the performances of these optical CCD data and the 2MASS 
photometry. This is crucial for future cluster studies as 2MASS becomes
widely used.

In the present work we explore M\,11 with 2MASS photometry.
In Sect.~2 the 2MASS photometry is presented. In Sect.~3
the cluster parameters are discussed. The cluster structure is
analyzed in Sect.~4. Luminosity and mass functions 
are discussed in Sect~5. Concluding remarks are given in Sect.~6.

\section{Database: 2MASS photometry}

\begin{figure}
 \resizebox{\hsize}{!}{\includegraphics{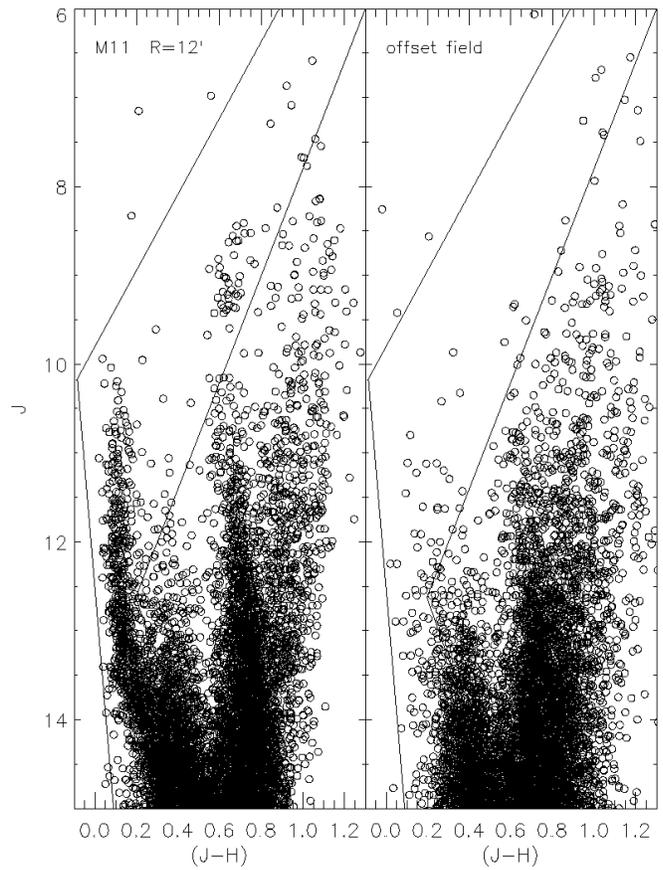}}
 \caption{$\rm J\times(J-H)$ CMDs of the central region of M\,11 and same-area background.
Cluster MS stars are identifiable as the leftmost sequence and cluster 
giant clump stars as the group around (J-H)=0.65 and J=9 in the central
region diagram (radius of 12'). Straight lines are CMD filters
used to exclude most of the field stars. The field background 
corresponds to a ring
with the same area of the central field with a maximum radius of 40'.}
\label{3378fg1}
\end{figure}

The 2MASS catalogue \citep{sss97} was employed in the 
present study because of the homogeneity
and the possibility of large-area data extractions. The near-infrared photometry
is also suitable for M\,11, since its MS (and giant clump) 
stands out from the
rich stellar field in CMDs as the one shown in Fig.~\ref{3378fg1}.  
A circular data extraction with radius $12\arcmin$ centered in M\,11 yielded 
8432 stars surmounting by 941 the number of stars in the background
field of same area (7491 stars), which is defined by an annulus with maximum 
radius of $40\arcmin$ (Fig.~\ref{3378fg1}). From J=10 to J=14 the cluster
MS seems to be little affected by field stars and should result in 
more precise determinations of the luminosity function (LF). A good 
account of the field 
is therefore necessary to obtain the LF of fainter stars. On this regard,
we advance that a statistical 
approach was employed in which the number of cluster stars in a magnitude 
interval is obtained from the difference between the total number of
stars within that interval at a given annulus and the same number at an
external annulus supposedly containing only field stars. 
Before this procedure the data are submitted to a CMD filter,
selecting only stars in the MS and giant cluster sequences 
(Fig.~\ref{3378fg1}).

The distance from the cluster center used to extract a fiducial 
background field was chosen
on the basis of a compromise between the premise that, on one side, the 
field is far enough in order to not contain cluster stars, and on the other
side, the field is close enough to keep at small levels the irregularities 
produced by dust and stellar density gradient.
Interesting to note that the field sequences correspond to disk stars
as shown by means of CMDs simulations \citep{mgc03}: the vertical
distribution around (J-H)=0.4 is formed by old disk turnoff stars
(m$\approx0.9$\,M$_{\odot}$), the one at (J-H)=0.75 is associated to 
old disk low MS (m$\leq0.6$\,M$_{\odot}$) (but also may include reddened 
giants) and the stars redder than (J-H)=0.9 are probably disk giants.

\subsection{Crowding effects}

M\,11 is a rich compact cluster, being classified as type ``II2r'' 
(detached, weak 
concentration; moderate range in brigthness; rich, more than 100 stars) 
by \cite{tru30}. Interestingly, \cite{rup66} has classified M\,11 as a globular
cluster of Trumpler type ``I2r'', even though its first colour-magnitude
diagram had already been obtained ten years before suggesting that the 
age of M\,11 is intermediate between that of the Pleiades and that of Praesepe
\citep{jsw56}. Not surprisingly, its concentration towards the center 
makes crowding an expected effect, enhanced by the instrumental 
limited spatial resolution. We have taken advantage of the complete 
analyses in the 2MASS database, 
which provide information on crowding for every source and band by means
of a flag (``cc\_flg''). This flag identifies whenever a source/band had
its photometry (flux) overestimated by at least 5\% due to
image artifacts, most of them  associated to crowded fields.

\begin{figure}
 \resizebox{\hsize}{!}{\includegraphics{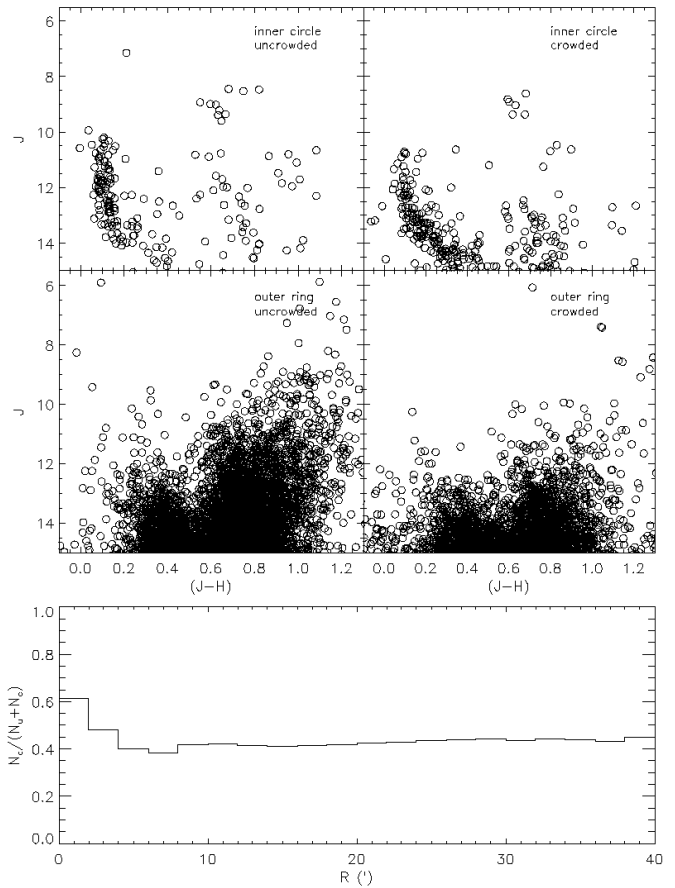}}
 \caption{Crowding evaluation over M\,11 concentric rings 2' wide.
CMDs of the inner circle and the outermost ring are shown for
the best quality data (uncrowded; left) and for data influenced by crowding
(right). Below is the distribution of the number of stars affected by crowding
relative to the total number of stars as a function of radius.}  
\label{3378fg2}
\end{figure}

We used this flag as an estimate of how significant is crowding
over the cluster radius.
The distribution of the ratio between the number of stars with 
photometry affected by crowding ($N_c$) and the total number of stars 
($N_u+N_c$) was calculated as a function of radius. 
The counts were carried out within rings 2' wide. The results are 
presented in Fig.~\ref{3378fg2}. A nearly constant distribution of 
$N_c/$($N_u+N_c$)
can be noticed except for the cluster inner regions (R$<4$'), in which 
crowding becomes important, as expected. What is the influence of crowding
on the LF? Since most stars affected by crowding in the cluster inner
regions follow the cluster sequences, i. e., the photometric precision is
not severely degraded for those stars, the LF should preserve its shape
if the magnitude bins are wider than the 
photometric uncertainty.
Magnitude bins of 0.5\,mag were used in the following since an uncertainty 
of 5\% in flux corresponds to 
$\sigma_{\rm J}=\frac{\sigma_{\rm F_J}}{\rm F_J}\frac{1}{0.4\ln{10}}\approx$0.05\,mag, 
about one tenth of the magnitude bin. To reach 0.5\,mag (bin width), the flux
should be overestimated by 50\%. Then, the crowding 
effects yield a negligible bias in the LF since large flux overestimates
seem not to be the case according to Fig.~\ref{3378fg2}, which shows that 
most of the stars affected by crowding are distributed over the same sequences
as those unaffected. Indeed, significant flux overestimates 
caused by crowding would be immediately detected in the CMDs of 
Fig.~\ref{3378fg2} by an overall smearing of the sequences.
Although completeness corrections were applied to
M\,11 by \cite{m84} and \cite{sbl99}, we have not applied such
procedure since we focused most of our analysis
outside the cluster core. 

\section{Cluster parameters}

The interstellar reddening towards M\,11 seems to be well established
(E(B-V)=$0.42\pm0.03$), with no evidence for a differential pattern across
the cluster field \citep[e. g.][]{sbl99}. The apparent 
distance modulus has been measured with different 
methods resulting $\rm 12.5<(V-M_V)<12.92$ \citep{sbl99,bcd93}. 
The cluster metallicity is nearly above solar, [Fe/H]=$0.136\pm0.086$, 
according to \cite{taa97}, which is approximately the value given in
WEBDA (Sect. 1).
 
\begin{figure}
 \resizebox{\hsize}{!}{\includegraphics{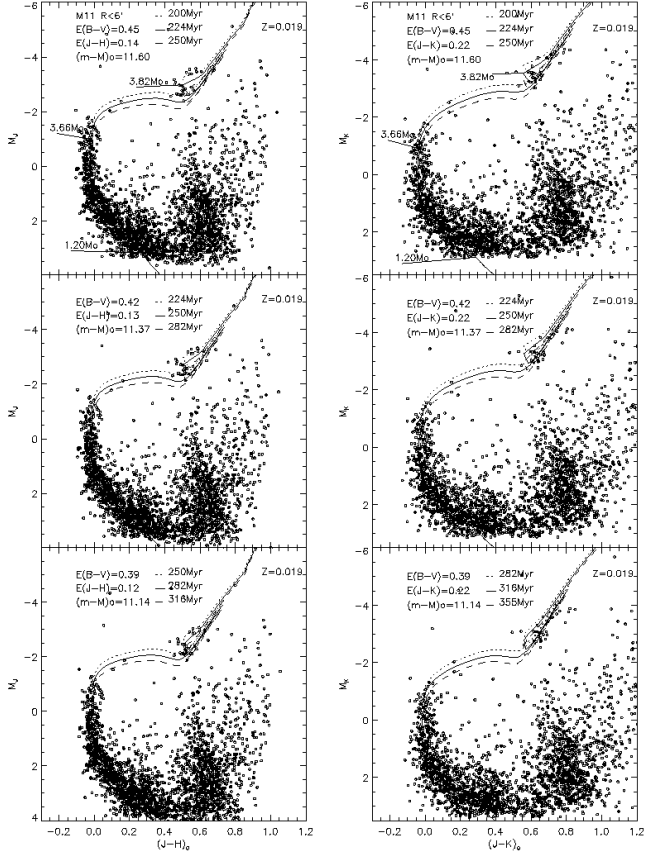}}
 \caption{CMDs of M\,11 central region and superimposed isochrones: 
M$_{\rm J}\times$(J-H)$_{\circ}$ left
and M$_{\rm K}\times$(J-K)$_{\circ}$ right. 
The parameter set employed to correct
the data are indicated as well as the solar-metallicity isochrone ages.
Top panels indicate selected stellar masses locus in the 224\,Myr isochrone.}  
\label{3378fg3}
\end{figure}

Cluster parameters were derived by fitting 
isochrones built using 2MASS filters \citep{bbg04}
to the cluster central region (R$<6$') CMDs J$\times$(J-H) and 
$\rm\ks\times(\jj-\ks)$, which best define the cluster sequences.
The cluster central region corresponds to its visual diameter and
it was chosen to maximize cluster members over field stars.
Isochrones were adjusted to both CMDs
using as constraints $\rm\ebv=0.42\pm0.03$ and 
(m-M)$_{\circ}=11.37\pm0.23$. Since each CMD was built from independent
observations involving a mixture of different bands, they provide different
data sets on which the isochrone matching should converge, giving more weight
to the analysis. Fig.~\ref{3378fg3} shows the best 
matching solar metallicity isochrones where the data have been 
corrected for the extreme and average values
that E(B-V) and (m-M)$_{\circ}$ may assume due to errors.
Selected stellar masses associated to the 224\,Myr isochrone 
are also indicated in the top panels: the lower mass (1.2\,M$_\odot$)
corresponds approximately to the data instrumental limit, the intermediate
mass (3.66\,M$_\odot$) locates the turnoff, and the higher mass 
(3.82\,M$_\odot$) marks the bluest point of the core He-burning phase (giant
clump).

By fixing E(B-V) and (m-M)$_{\circ}$ according to observational
constraints, the free parameters were reduced to the isochrone age 
and metallicity. 
The well-defined main sequence (MS) and giant clump of M\,11 allow an 
unambiguous derivation of its age with an uncertainty of nearly 10\%. 

A good overall match is obtained if the average reddening 
and average true distance modulus are used for the 250\,Myr isochrone 
(Fig. \ref{3378fg3}, middle panels), with satisfactory results also being
obtained for the 224\,Myr isochrone and the 282\,Myr one.
In spite of this, a difference of $\approx0.05$mag. is found for the 
giant clump mean locus in the M$_{\rm J}\times$(J-H)$_{\circ}$ CMD. 

\begin{figure}
 \resizebox{\hsize}{!}{\includegraphics{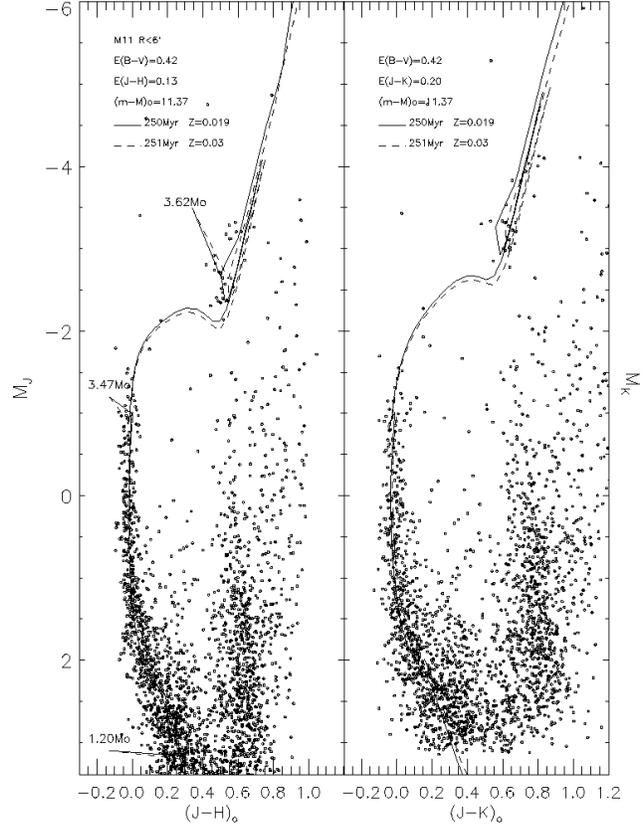}}
 \caption{CMDs of M\,11 central region with isochrones superimposed: 
M$_{\rm J}\times$(J-H)$_{\circ}$ left
and M$_{\rm K}\times$(J-K)$_{\circ}$ right. The parameter set employed to correct
the data are indicated together with two matched isochrones of similar age and 
different metallicities.
Selected stellar mass loci are indicated for both isochrones.}  
\label{3378fg4}
\end{figure}

Taking into account the fact that the cluster has metallicity 
above solar \citep{taa97},  the best match isochrone
(250\,Myr) is presented in Fig.~\ref{3378fg4} for two metallicities 
(Z=0.019 and Z=0.03
or [Fe/H]=0.20) together with the corrected data in both CMDs. 
Three mass values are shown connected to the corresponding isochrone.
This comparison indicates that the colour of the cluster stars in the 
clump is not
due to a cluster metallicity higher than solar, indeed the higher 
metallicity isochrone indicates a redder colour for the clump. Binaries may be
affecting the cluster clump colour, since they would explain a brightening
of clump stars if they were in binary systems and a blueing if they 
comprise a red giant and a blue turnoff star.   

In conclusion, an
age of t$=250\pm30$Myr was obtained for M\,11 and the solar metallicity 
isochrone was employed as representative of the cluster stellar population.
Such a representation is relevant in the determination of the mass 
function since the isochrone is the source of the mass-luminosity relation
used to transform the observed luminosity function (LF) into the mass 
function (MF). Indeed, as it is well known, the precise location of 
masses over the MS and the clump are 
influenced by age and metallicity as can be noticed by comparing 
Fig.~\ref{3378fg3} and Fig.~\ref{3378fg4}. Thus, the mass range of 
observed stars and, in consequence, the cluster MF are partially determined
by the isochrone chosen. 

\section{Cluster structure from the King-profile}

\begin{figure}
 \resizebox{\hsize}{!}{\includegraphics{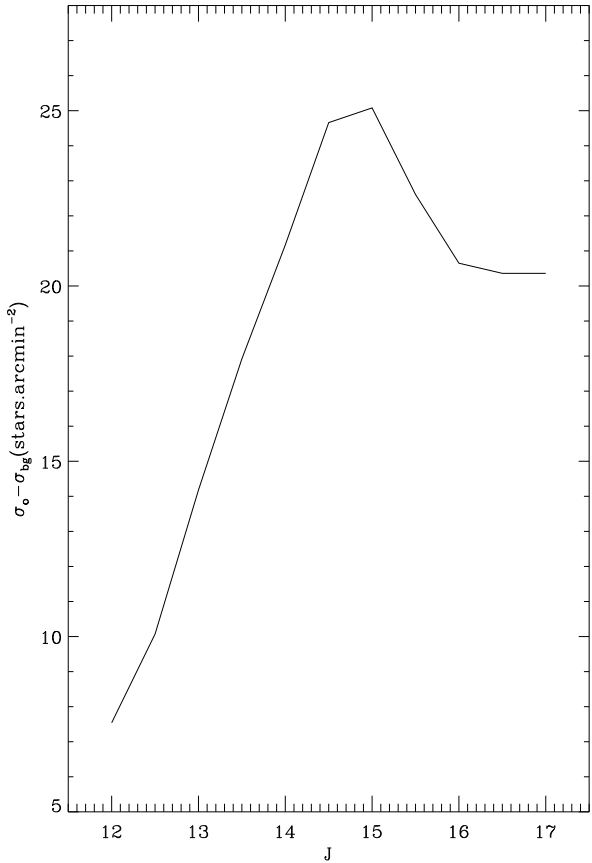}}
 \caption{Central surface density minus background surface density as a 
function of J cutoff magnitude. The curve peak reveals the J cutoff at
which there is an optimal separation between cluster stars and field stars.}
\label{3378fg5}
\end{figure}

The colour-magnitude filter (see Fig.~\ref{3378fg1}) in the plane 
J$\times$(J-H) was applied in order 
to select the CMD regions
containing the cluster evolutionary sequences.
The magnitude cutoff at the lower MS end adopted for fitting a King-profile 
is based on 
the optimal separation of cluster stars and background field. Fig.~\ref{3378fg5}
shows that at J=15.0 the density of cluster stars with respect to the
background reaches a maximum value of 25.08\,stars.arcmin$^{-2}$ at the central
circle of 1\,arcmin of radius. Therefore, the magnitude cutoff at J=15.0
was adopted.

\begin{figure}
 \resizebox{\hsize}{!}{\includegraphics{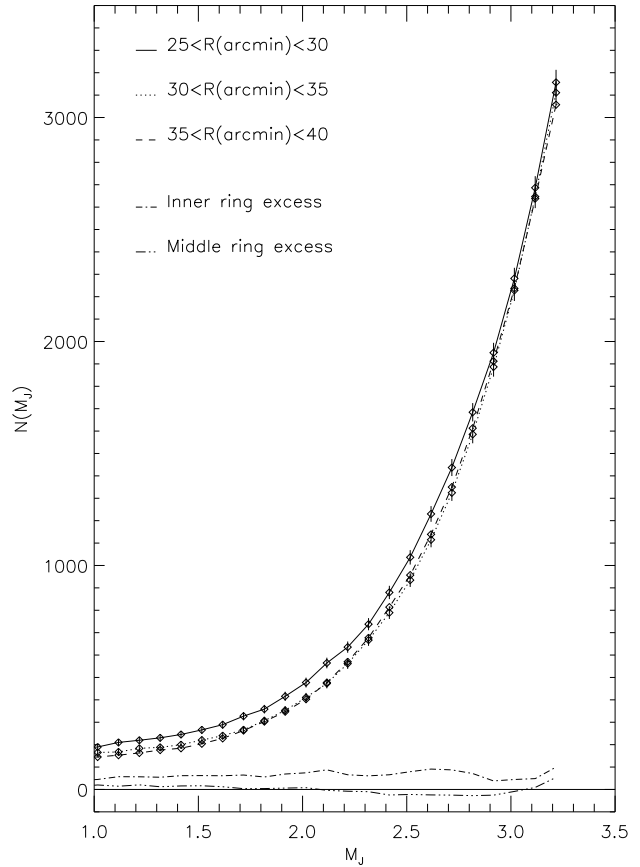}}
 \caption{The cumulative LFs for three $5\arcmin$-wide outer rings.
The counts are normalized to the outer ring area. The curve excesses 
indicated by ``inner ring excess'' and ``middle ring excess'' 
are differences between a given ring LF and the outer ring one.}
\label{3378fg6}
\end{figure}

A discussion on the effects of applying colour-magnitude filters
and background selection are given in \cite{bbs04}. Because the background 
selection is critical for M\,11, the filtered cumulative 
distribution of stars in the cluster outer regions was analyzed
as a function of M$_{\rm J}$. In Fig.~\ref{3378fg6} the cumulative
LF of the three outer rings 5\,arcmin wide are compared. The counts were
carried out within bins of $\Delta({\rm M}_{\rm J})=0.1$\,mag and 
the range of M$_{\rm J}$
presented in this Figure corresponds to the cluster MS. The LFs are normalized
to the area of the outer ring ($35<$R(')$<40$). A clear excess 
in the cumulative LF is noticed for the inner ring ($25<$R(')$<30$),
which is better visualized by the difference between its cumulative LF
and that for the outer ring, and characterized as the ``inner ring excess'' 
in Fig.~\ref{3378fg6}. The middle ring ($30<$R(')$<35$) excess is
also shown, which indeed does not reveal any significant 
difference between the middle and outer cumulative LFs, presumably for
representing both a fiducial background field, little affected by cluster
stars. In contrast, the inner ring clearly reveals the presence 
of cluster stars. Therefore cluster stars are present and dominant
over the field for distances less than 30' from its center. 

\begin{figure}
 \resizebox{\hsize}{!}{\includegraphics{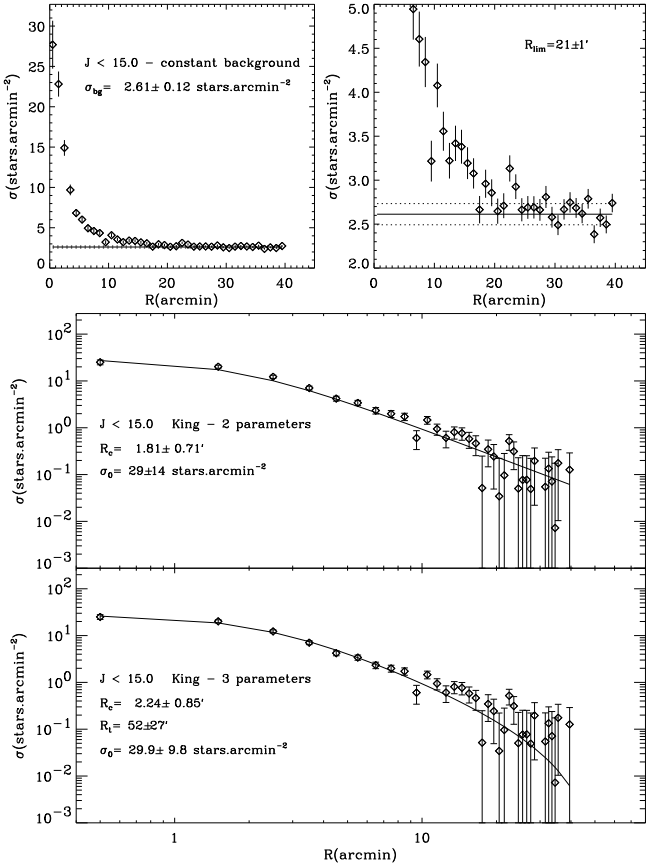}}
 \caption{Radial profile of surface stellar density for a cutoff at 
J=15.0. Top-left: the overall
profile with a fitted constant background ($\sigma_{\rm bg}$) superimposed;
Only stellar densities at $R>$30' were considered in the fit. 
Top-right: zoom 
of the radial profile showing more clearly the fitted background (continuous 
line) bounded by the corresponding 1-$\sigma$ dispersion (dotted lines).
Middle: 2-parameters (central stellar surface density and core radius) 
King-model fitted to the radial profile in log scales.
Bottom: 3-parameters (central stellar surface density, core radius and tidal
radius) King-model fitted to the radial profile in log scales. The best-fit
parameter values are indicated. Error bars denote 1-$\sigma$ poissonian 
fluctuations.}
\label{3378fg7}
\end{figure}

An analysis of the cluster structure was performed on this selected 
sample. The radial distribution of stellar surface density 
(stars/arcmin$^2$) was investigated by counting stars within rings 
of width 1 up to 40' from the cluster center. As expected,
the procedure carried out to maximize cluster stars over field stars 
helps to enhance the cluster structure keeping the background field at
acceptable levels, as shown in the top panels of Fig.~\ref{3378fg7}.
In this Figure, a constant background was fitted to the outer region 
sampled, $30<$R(')$<40$, its 1-$\sigma$ dispersion being shown in the
top-right panel.
 
A reasonable limiting radius for M\,11 is R$_{lim}=21\pm1$\,', where the
cluster star density begins to stand out from the background one. 
The fitted constant background was then subtracted from the overall
surface density and a King-profile fitting was performed.
Two-parameter (central stellar surface density, $\sigma_0$, and core radius,
R$_{\rm c}$) and three-parameter ($\sigma_0$, R$_{\rm c}$ and tidal radius,
R$_{\rm t}$) King functions \citep{k62,k66} were employed. The fitted 
functions are 
presented in Fig~\ref{3378fg7} in log scales (bottom) together with 
the best fit parameters. 
The two-parametric King function should better represent the cluster inner
regions and the three-parametric King function should provide a better estimate
of the cluster overall structure. In both fittings the estimates of inner 
parameters ($\sigma_0$ and R$_{\rm c}$) agree within the uncertainties. 
The tidal radius was estimated with 50\% precision because of the 
sensitivity of the three-parameter King model to the fluctuations in the
density of cluster stars in its outskirts, almost at the background level.
Such fluctuations (represented by poissonian errors in Fig.~\ref{3378fg7}) 
are taken into account in the fitting by applying a weigthed 
least-squares method.   

The adopted true distance modulus (m-M)$_{\circ}=11.37\pm0.23$
translates into a distance from the Sun of d$_{\odot}=1.89\pm0.26$\,kpc. Therefore, the 
linear limiting radius of M\,11 is  R$_{lim}=11.5\pm1.7$\,pc 
(Fig.~\ref{3378fg7}). A
galactocentric distance of d$_{\rm GC}=6.38\pm0.21$\,kpc is obtained using
d$_{GC}=8.0$\,kpc for the Sun Galactocentric distance \citep{r93}.

\begin{figure}
 \resizebox{\hsize}{!}{\includegraphics{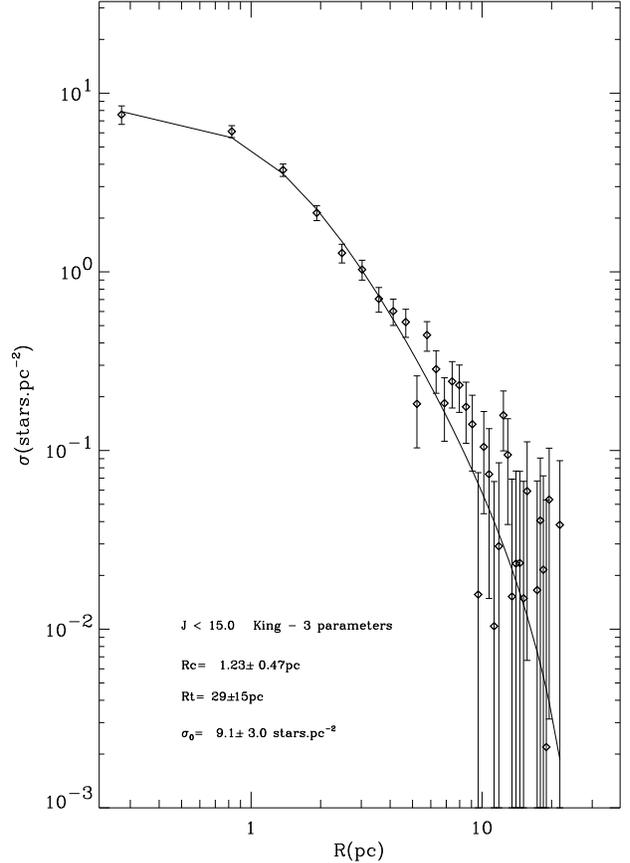}}
 \caption{Same as the lower panel of Fig.~\ref{3378fg7} but in physical
units.}
\label{3378fg8}
\end{figure}

In Fig.~\ref{3378fg8} the same three-parametric King function shown
in  Fig.~\ref{3378fg7} is presented in absolute units, where 
1\,arcmin=0.55\,pc. The cluster structural parameters concerning
stars with J$\leq15.0$ are 
$\sigma_0=9.1\pm3.0$\,stars.pc$^{-2}$, R$_{\rm c}=1.23\pm0.47$\,pc and
R$_{\rm t}=29\pm15$\,pc. 
The core radius is 1.7 times larger than that quoted by \cite{nspm02}.

A deviation from the King profile can be seen 
between 6 and 9\,pc (which is inside R$_{lim}$, but well beyond 
R$_{\rm c}$) where the cluster star 
density is in excess with respect 
to the model (Fig.~\ref{3378fg8}). Such an excess is expected 
if the cluster is in the
process of loosing low mass stars by means of energy equipartition.
If so, this excess of stellar surface density is also expected 
in the cluster outskirts, but detecting this effect
beyond 9\,pc is more difficult because of the uncertainties 
produced by the background field.

\begin{figure}
 \resizebox{\hsize}{!}{\includegraphics{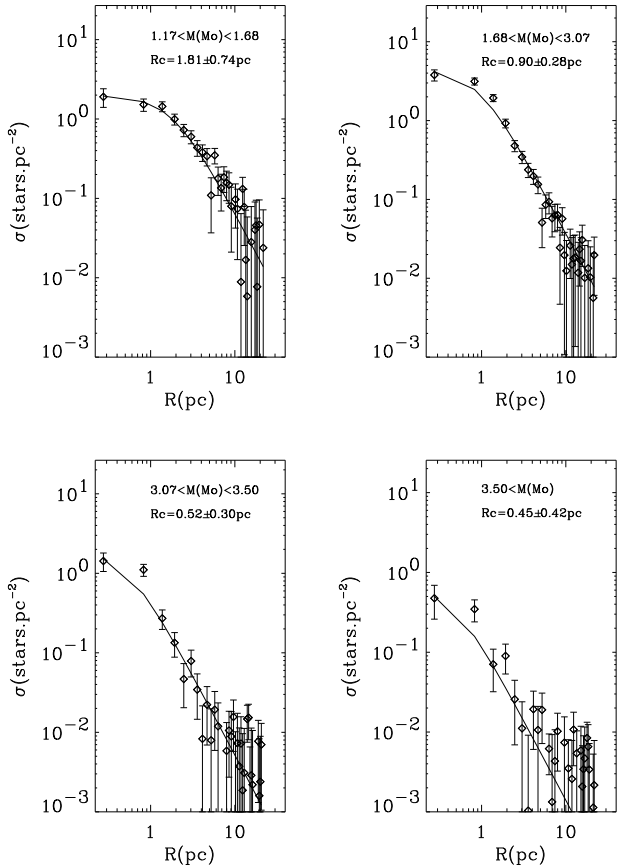}}
 \caption{Radial profile of surface density for 4 stellar mass ranges. 
The best-fit two-parameter King models are plotted and the corresponding
core radius indicated.}
\label{3378fg9}
\end{figure}

Alternative means of diagnosing mass segregation in clusters have been
successfully employed by e.g. \cite{rm98a}, \cite{rm98b}, \cite{m84}. 
Specifically,
following these studies, we derive the cluster structure and cumulative 
distributions as characterized
by stars in four mass ranges: $1.17<$m$(\ms)<1.68$ (lower MS, 
$2.00<$M$_{\rm J}<3.27$),
$1.68<$m$(\ms)<3.07$ (intermediate MS, $0.00<$M$_{\rm J}<2.00$), 
$3.07<$m$(\ms)<3.50$ (upper MS, $-2.00<$M$_{\rm J}<0.00$)
and m$(\ms)>3.50$ (giants, M$_{\rm J}<-2.00$). Two-parametric King 
models were fitted
to the radial stellar density profile for each mass range separately.
The results are presented in Fig.~\ref{3378fg9}, where the best solution
for the core radius is shown. The more extended populations (larger 
R$_{\rm c}$) correspond to those populations composed by less massive 
stars, as one would expect for
a dynamically evolved cluster with conspicuous mass segregation. Although
satisfactory fittings were obtained for the lower and intermediate MS,
reasonable fittings were not achieved for the upper MS and giants, whose
less numerous samples are subject to larger statistical errors. Therefore 
R$_{\rm c}$ should be taken carefully as an indicator of
mass segregation in the cluster. 

\begin{figure}
 \resizebox{\hsize}{!}{\includegraphics{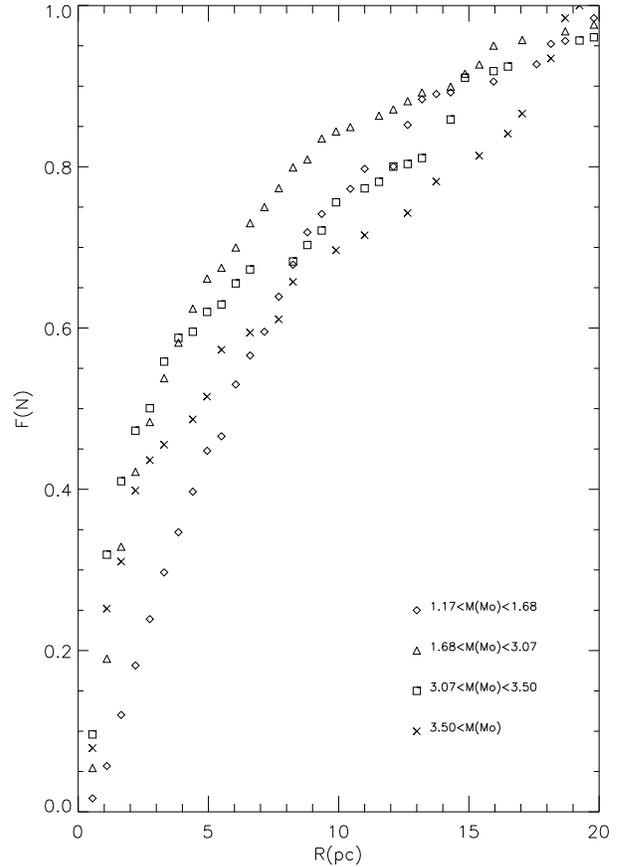}}
 \caption{Cumulative distributions of stars in 4 mass ranges corresponding
to the lower MS (diamonds), intermediate MS (triangles), upper MS (squares)
and giants (crosses).}
\label{3378fg10}
\end{figure}

Indeed, a better, direct account of 
mass segregation is revealed by the cumulative distribution of stars 
in the mass ranges as above (Fig.~\ref{3378fg10}). The cumulative 
distribution (F(N)), field subtracted and determined up to 20\,pc 
($\approx$ 2$\times$R$_{lim}$), shows that less stars in the
lower MS sample are concentrated in the inner 6-7\,pc in comparison
with the other more massive samples. Closer to the cluster center,
at R$<3$\,pc, the number of stars more massive than 1.68\,$\ms$ 
increases more sharply than the number of lower mass stars, indicating
mass segregation. Comparing the lower MS with the intermediate MS,
the former is less concentrated, in accord with the mass segregation 
effects. In the outer regions at R$>10$\,pc, the number of 
giant stars appears to be still increasing, a feature already 
noticed by \cite{m84}. \cite{m84} consider mass loss as 
a possible explanation for this unexpected effect: by losing 
mass the giant stars would reproduce the distribution of lower mass stars.  

For being at the present time relatively close to the Galactic center,
M\,11 may have suffered the consequences
of strong interactions with molecular clouds and/or tidal effects of the 
Galaxy determining its 
structure by affecting its tidal radius. In this respect, the internal
dynamical processes occurring in M\,11 are less effective in shaping
the cluster overall structure than in the case of the older cluster
NGC\,188, which orbits
the Galaxy in a path beyond the solar circle \citep{bbs04}.  An orbit
calculation for M\,11 would help to constrain its dynamical properties.

\section{Radial luminosity and mass functions}

\begin{figure}
 \resizebox{\hsize}{!}{\includegraphics{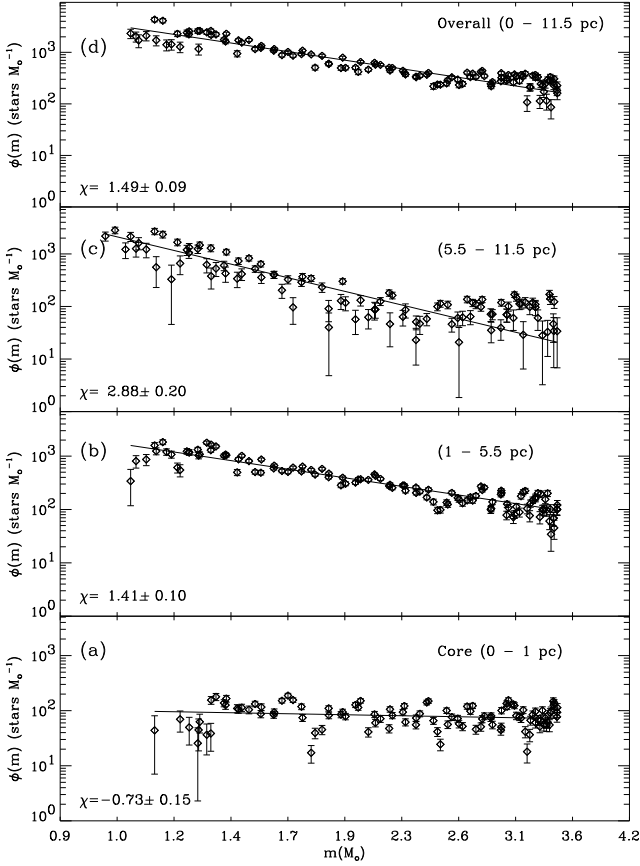}}
 \caption{Mass functions of (a)~core, (b)~intermediate region,
(c)~external corona and (d)~the overall MF are shown with power-law
fittings resulting the slopes indicated. The regions are indicated 
by the radius of annular extractions. The error bars correspond to
1-$\sigma$ fluctuations.}
\label{3378fg11}
\end{figure}

LFs of the filtered CMD were determined for each 2MASS band separately 
and different annular 
regions corresponding to the cluster
core, an intermediate annulus (R$_{\rm c}<$R$<5.5$\,pc), the halo 
(5.5\,pc$<R<$R$_{\rm lim}$) and the overall cluster extension 
(0\,pc$<$R$<$R$_{\rm lim}$). The LFs have been properly corrected for 
the background field by subtracting the counts measured in the outer 
field ($30<$R(')$<40$) from the counts per magnitude bin in
each region scaled to their area.
The main-sequence LFs constructed independently for each band 
were converted
in one MF for each cluster region by fitting mass-luminosity relations
(in J, H and \ks) from the solar metallicity 250\,Myr isochrone.   
The overall MF and the MFs of the selected spatial regions are presented
in Fig.~\ref{3378fg11} together with power-law MF 
($\phi({\rm m})={\rm A}.{\rm m}^{-(1+\chi)}$, where A is the MF normalization
and $\chi$ is the MF slope) weighted fittings. The resulting MF slopes are
shown in Table~\ref{tab2}. The regions sampled are the core, the inner and
outer halo and the overall cluster. The MF slope in the core is very flat, 
comparable to those of the cores of M\,93, NGC\,2477 and NGC\,3680 
\citep{bb05}. M\,11 presents a MF slope gradient
from the core to the outer regions, which is characteristic of large-scale
mass segregation. The overall MF slope value is similar to the standard
Salpeter one, and comparable to most of the classical open clusters
studied in \cite{bb05}. 
 
In order to determine the total cluster mass from the overall MF, it was 
extrapolated to m$<1.0$\,M$_{\odot}$ down to the H-burning limit using the 
universal MF model by \cite{k01}, which gives 
$\chi=1.3\pm0.3$ for $0.5<$m(M$_{\odot}$)$<1.0$ and $\chi=0.3\pm0.5$ for 
$0.08<$m(M$_{\odot}$)$<0.5$.
The slope $\chi=1.49\pm0.09$ was applied to the observed portion of the MF,
which spans MS masses between $1.09<$m(M$_{\odot}$)$<3.45$.
Following this procedure we get m(MS)=($10.8\pm3.8$)$\times10^3$\,M$_{\odot}$.

The mass content in evolved stars was computed from the integrated LF
above the turnoff at M$_{\rm J}=-2.0$ and the corresponding 
isochrone mass interval of the CMD observed stars.
According to the 250\,Myr isochrone, we observe evolved stars in 
the mass range $3.50<$m(M$_\odot$)$<3.74$
with a strong concentration around m=3.62\,M$_{\odot}$ in the giant 
clump region (see Fig.~\ref{3378fg4}). We assume a typical mass for the
evolved stars in M\,11 of m$=3.6\pm0.1$\,M$_{\odot}$ and use it to get
the total mass in evolved stars, i. e., m(post-MS)=$176\pm26$\,M$_{\odot}$.

Therefore, the total mass obtained for M\,11 inside a circular area 
of radius 11.5\,pc is ($11.0\pm3.8$)$\times10^3$\,M$_{\odot}$.

MFs in the selected regions of the cluster were used to compute
the total mass of the subsystems. They are presented in Table~1.

\begin{table*}
\caption[]{Data derived from the mass functions in M\,11.}
\begin{tiny}
\label{tab2}
\renewcommand{\tabcolsep}{0.72mm}
\begin{tabular}{cccccccccccccccccc}
\hline\hline
&\multicolumn{2}{c}{Evolved}&&\multicolumn{1}{c}{$\chi$}&&\multicolumn{4}{c}{$\rm
Observed+Evolved$}&&\multicolumn{4}{c}{$\rm Extrapolated+Evolved$}\\
\cline{2-3}\cline{5-5}\cline{7-10}\cline{12-15}\\
Region&N$^*$&\mevol&&$1.09-3.45$&&$\rm
N^*$&\mobs&$\sigma$&$\rho$&&$\rm N^*$&\mtot&$\sigma$&$\rho$\\
(pc)&(Stars)&($10^2\ms$)&& &&($10^3$stars)&($10^3\ms$)&($\rm
\ms\,pc^{-2}$)&($\rm \ms\,pc^{-3}$)&& ($10^3$stars)&
($10^3\ms$)&($\rm \ms\,pc^{-2}$)&($\rm \ms\,pc^{-3}$)\\
 (1)&  (2) & (3)   && (4)     &&(5)  &(6) & (7)  & (8)  &&
(9)&(10)&(11)&(12)\\
\hline
$0.0-1.0$&$16\pm4$&$0.6\pm0.1$&&$-0.73\pm0.15$&&$0.20\pm0.03$&$0.48\pm0.08$&$153\pm24$&$115\pm18$&&$0.34\pm0.04$&$0.56\pm0.08$&$177\pm26$&$133\pm19$&\\

$1.0-5.5$&$25\pm5$&$0.9\pm0.2$&&$1.41\pm0.10$ &&$1.01\pm0.08$&$1.82\pm0.15$&$19.8\pm1.6$&$2.63\pm0.22$&&$15\pm11$&$5.9\pm2.0$&$64\pm22$&$8.5\pm2.9$&\\

$5.5-11.5$&$16\pm4$&$0.6\pm0.1$&&$2.88\pm0.20$&&$0.84\pm0.08$&$1.24\pm0.13$&$3.87\pm0.41$&$0.22\pm0.02$&&$19\pm14$&$6.4\pm2.6$&$20\pm8$&$1.1\pm0.5$&\\

$0.0-11.5$&$49\pm7$&$1.8\pm0.3$&&$1.49\pm0.09$&&$1.84\pm0.13$&$3.29\pm0.30$&$7.93\pm0.59$&$0.52\pm0.04$&&$28\pm20$&$11.0\pm3.8$&$26\pm9$&$1.7\pm0.6$&\\

\hline\hline
\end{tabular}
\begin{list}{Table Notes.}
\item Col.~4 gives the MF slope derived for the observed MS mass range. 
The mass of the evolved  stars is included in \mobs\
(col.~6) and \mtot\ (col.~10).
\end{list}
\end{tiny}
\end{table*}

The relation between the tidal radius and the cluster mass as given by
\cite{k62} was also used to compute a total mass for M\,11. Mathematically:
m=3M$_{\rm G}(\frac{{\rm R}_{\rm t}}{{\rm R}_{\rm p}})^3$,
where R$_{\rm p}$ is the perigalacticon distance and M$_{\rm G}$ is
the Galactic mass inside R$_{\rm p}$. Assuming a nearly circular orbit
for M\,11, and consequently R$_{\rm p}$=d$_{\rm CG}$=$6.38\pm0.21$\,kpc,
and M$_{\rm G}=8.0\times10^{10}$M$_{\odot}$ \citep{cc94}, we get an expected
mass of $\rm m=(22\pm35)\times10^3\,\ms$, the large error coming from the uncertainty
in the tidal radius. Such a theoretical value is about 2 times larger
than our estimate.

The relaxation time of a star system can be defined as 
t$_{\rm relax}=\frac{\rm N}{\rm 8lnN}$t$_{\rm cross}$, where 
t$_{\rm cross}=$R/$\sigma_{\rm v}$ is the crossing time, N is the
total number of stars and $\sigma_{\rm v}$ is the velocity dispersion
\citep{bt87}. t$_{\rm relax}$ is the characteristic time scale in which
the cluster reaches some level of kinetic energy equipartition with
massive stars sinking to the core and low-mass stars being transferred
to the halo. Using the velocity dispersion found for M\,11 of 
$\sigma_{\rm v}=2.9$\,km/s \citep{ms77} we obtain 
t$_{\rm relax}\approx1300$\,Myr for the whole cluster and
t$_{\rm relax}\approx2$\,Myr for the cluster core.
The MF slope flattening towards the center, as an evidence 
of mass segregation observed in M\,11 is consistent with the cluster
core being dynamically evolved in agreement with t$_{\rm relax}$ in the 
core, which is smaller than the cluster age.

\section{Concluding remarks}

We employed 2MASS photometry to explore the structure and stellar content of
the open cluster M\,11, which is located internal to the Solar circle. The
near-IR photometry basically confirmed previous photometric parameters
derived from the optical. We studied this cluster with spatial resolution, 
owing to the wide-angle analysis allowed by 2MASS data. We obtained 
a core radius 
of 1.23\,pc and a tidal radius of 29\,pc. The latter value was obtained by 
fitting the three-parameter King profile to the radial distribution of stars,
which is possible only for the more populous and highest-contrast open
clusters.

The spatial distribution of mass functions showed a very flat one in the core 
($\chi=-0.73$) and a steep halo one ($\chi=2.88$), particularly in the outer region. 
Mass segregation is implied by the results. The overall mass function slope is similar 
to a standard Salpeter one. The spatial distribution of
mass function slopes derived from 2MASS agrees with that derived from 
optical CCD data, which further confirms the reliability of 2MASS data for future analyses 
of this kind at comparable observational limits. 

The cluster is massive, with a total (extrapolating the mass function to 
0.08\,M$_\odot$)
mass of $\approx11\,000\,\ms$, which is somewhat larger than previous estimates.

The large cluster mass of M\,11 is a slowing down factor of dynamical evolution
because of a longer relaxation time.
However, its position well within the Solar circle is expected to speed it up
because of stronger tidal effects 
(e.g. \citealt{delaF2002}; \citealt{Bergond2001}; 
\citealt{bb05}).
 
\begin{acknowledgements}
This publication makes use of data products from the Two Micron All Sky Survey, which 
is a joint project of the University of Massachusetts and the Infrared Processing and 
Analysis Center/California Institute of Technology, funded by the National Aeronautics 
and Space Administration and the National Science Foundation. We also 
thank the referee, Dr. J.-C. Mermilliod, for helping to improve the
work and for the use of the WEBDA open cluster database. We acknowledge 
support from the Brazilian 
Institutions CNPq and FAPEMIG.

\end{acknowledgements}

\bibliographystyle{apj}
\end{document}